\documentclass[11pt,twoside]{article}
\usepackage{asp2010}

\resetcounters

\usepackage[english]{babel}
\usepackage{graphicx}
\usepackage{bm}
\usepackage{booktabs}
\usepackage{color}
\usepackage{hyperref}
\usepackage{latexsym}
\usepackage{natbib}
\usepackage{times}
\usepackage{stmaryrd}

\newcommand{\ie}{\textit{i.e.}~}

\newcommand{\st}{\textit{s.t.}~}

\begin{document}

\title{Universality and intermittency in relativistic turbulent flows of
a hot gas}
\author{David Radice$^1$ and Luciano Rezzolla$^{1,2}$
\affil{$^1$Max-Planck-Institut f\"ur Gravitationsphysik, Albert Einstein
Institut, Potsdam, Germany}
\affil{$^2$Department of Physics and Astronomy, Louisiana State
University, Baton Rouge, USA}}

\begin{abstract}
  With the aim of determining the statistical properties of
  relativistic turbulence and unveiling novel and non-classical
  features, we present the results of direct numerical simulations of
  driven turbulence in an ultrarelativistic hot plasma using
  high-order numerical schemes. We study the statistical properties of
  flows with average Mach number ranging from $\sim 0.4$ to $\sim 1.7$
  and with average Lorentz factors up to $\sim 1.7$. We find that flow
  quantities, such as the energy density or the local Lorentz factor,
  show large spatial variance even in the subsonic case as
  compressibility is enhanced by relativistic effects. The velocity
  field is highly intermittent, but its power-spectrum is found to be
  in good agreement with the predictions of the classical theory of
  Kolmogorov.
\end{abstract}
\section{Introduction}
Turbulence is an ubiquitous phenomenon in nature as it plays a
fundamental role in shaping the dynamics of systems ranging from the
mixture of air and oil in a car engine, up to the rarefied hot plasma
composing the intergalactic medium. Relativistic hydrodynamics is a
fundamental ingredient in the modeling of a number of systems
characterized by high Lorentz-factor flows, strong gravity or
relativistic temperatures. Examples include the early Universe,
relativistic jets, gamma-ray-bursts (GRBs), relativistic heavy-ion
collisions and core-collapse supernovae \citep{Font08}.

Despite the importance of relativistic hydrodynamics and the
reasonable expectation that turbulence is likely to play an important
role in many of the systems mentioned above, extremely little is known
about turbulence in a relativistic regime. For this reason, the study
of relativistic turbulence may be of fundamental importance to develop
a quantitative description of many astrophysical systems. To this aim,
we have performed a series of high-order direct numerical simulations
of driven relativistic turbulence of a hot plasma.
\section{Model and method}
We consider an idealized model of an ultrarelativistic fluid with
four-velocity $u^{\mu} = W (1, v^i)$, where $W \equiv (1 -
v_iv^i)^{-1/2}$ is the Lorentz factor and $v^i$ is the three-velocity
in units where $c = 1$. The fluid is modeled as perfect and described
by the stress-energy tensor 
\begin{equation}
T_{\mu\nu} = (\rho + p) u_{\mu} u_{\nu} + p\, g_{\mu\nu}\,,
\end{equation}
where $\rho$ is the (local-rest-frame) energy density,
$p$ is the pressure, $u_{\mu}$ the four-velocity, and $g_{\mu\nu}$ is
the spacetime metric, which we take to be the Minkowski one. We evolve
the equations describing conservation of energy and momentum in the
presence of an externally imposed Minkowskian force $F^{\mu}$, \ie
\begin{equation}
\nabla_{\nu} T^{\mu\nu} = F^{\mu},
\end{equation}
where the forcing term is written as $F^{\mu} = \tilde{F}(0, f^i)$.  More
specifically, the spatial part of the force, $f^i$, is a zero-average,
solenoidal, random, vector field with a spectral distribution which has
compact support in the low wavenumber part of the Fourier spectrum.
Moreover, $f^{i}$, is kept fixed during the evolution and it is the same
for all the models, while $\tilde{F}$ is either a constant or a simple
function of time (see \cite{Radice2012b} for details).

\begin{figure*}
  \begin{minipage}{0.49\hsize}
    \includegraphics[width=\columnwidth]{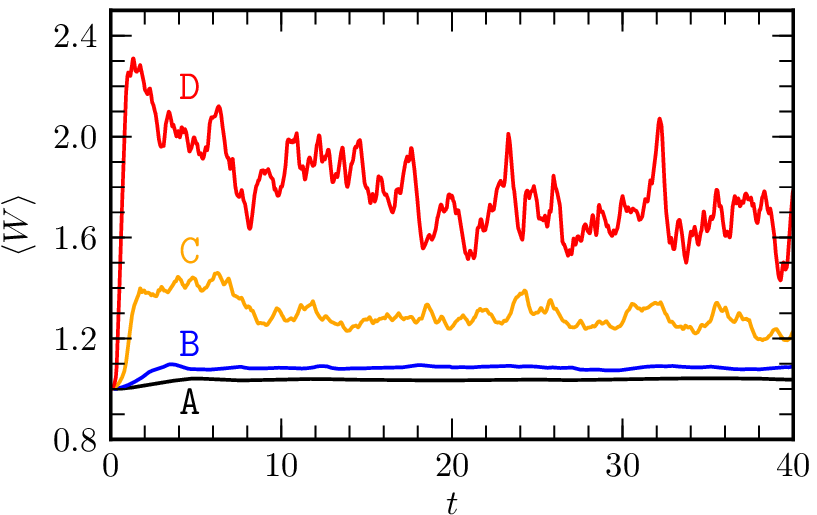}
  \end{minipage}
  \begin{minipage}{0.49\hsize}
  \begin{center}
    \includegraphics[width=0.8\columnwidth]{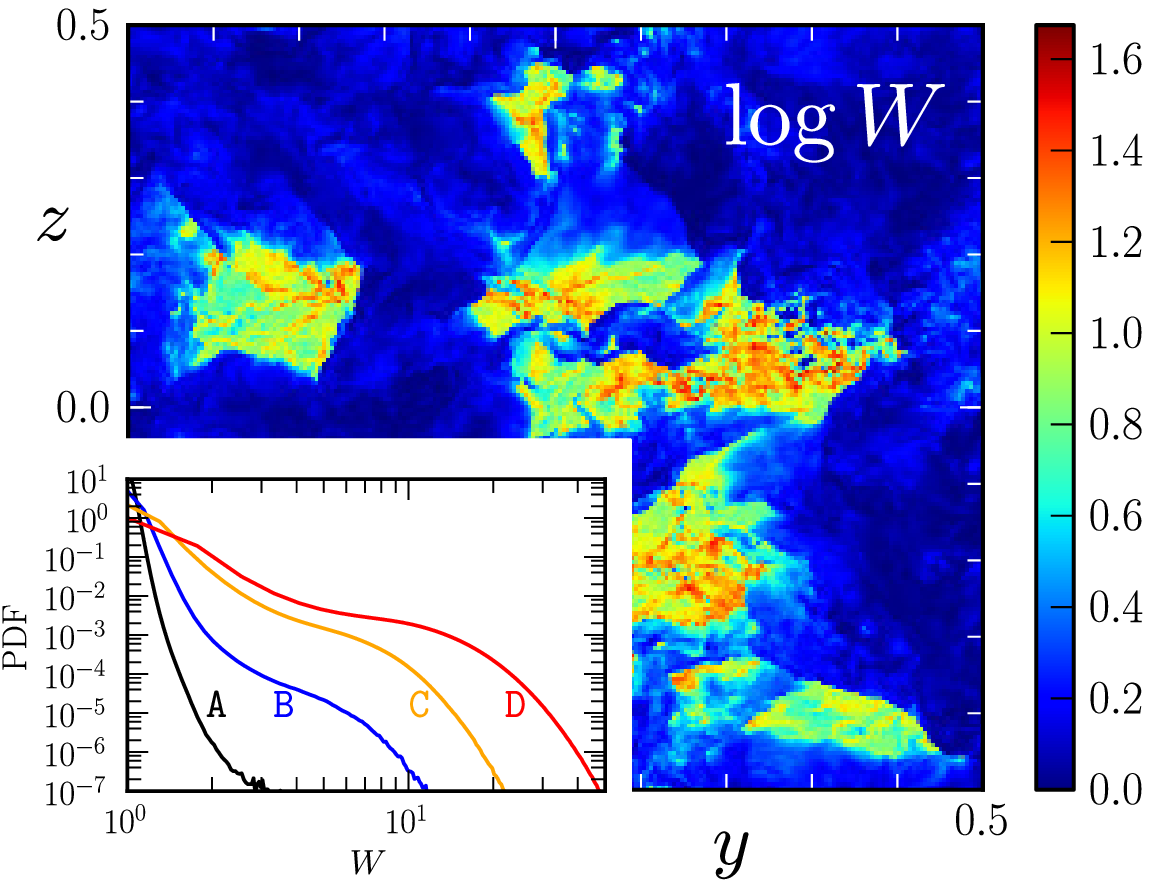}
  \end{center}
  \end{minipage}
  \caption{
  \textit{Left panel:} average Lorentz factor as a function of time for
  the different models considered. Note that a quasi-stationary state is
  reached before $t\sim 10$ for all values of the driving force.
  \textit{Right panel:} logarithm of the Lorentz factor on the $(y,z)$
  plane at the final time of model \texttt{D}. Note the large spatial
  variations of the Lorentz factor with front-like structures. The
  time-averaged PDFs are shown in the lower left corner for the different
  models considered.
  \label{fig:lorentz}
  }
\end{figure*}

The set of relativistic-hydrodynamic equations is closed by the
equation of state (EOS) $p = \frac{1}{3} \rho$, thus modelling a hot,
optically-thick, radiation-pressure dominated plasma, such as the
electron-positron plasma in a GRB fireball or the matter in the
radiation-dominated era of the early Universe. The EOS used can be
thought as the relativistic equivalent of the classical isothermal EOS
in that the sound speed is a constant, \ie $c_s^2 = 1/3$. At the same
time, an ultrarelativistic fluid is fundamentally different from a
classical isothermal fluid. For instance, its ``inertia'' is entirely
determined by the temperature and the notion of rest-mass density is
lost since the latter is minute (or zero for a pure photon gas) when
compared with the internal one. For these reasons, there is no direct
classical counterpart of an ultrarelativistic fluid and a relativistic
description is needed even for small velocities.

We solve the equations of relativistic hydrodynamics in a 3D periodic
domain using the high-resolution shock capturing scheme described
in \citep{Radice2012a}. In particular, ours is a flux-vector-splitting
scheme \citep{Toro99}, using the fifth-order MP5
reconstruction \citep{suresh_1997_amp}, in local characteristic
variables \citep{Hawke2001}, with a linearized flux-split algorithm
with entropy and carbuncle fix\\ \citep{Radice2012a}.
\section{Basic flow properties}
Our analysis is based on the study of four different models, which we
label as \texttt{A}, \texttt{B}, \texttt{C} and \texttt{D}, and which
differ for the initial amplitude of the driving factor $\tilde{F}=1, 2,
5$ for models \texttt{A}--\texttt{C}, and $\tilde{F}(t) = 10 +
\frac{1}{2} t$ for the extreme model \texttt{D}. Each model was evolved
using three different uniform resolutions of $128^3$, $256^3$ and $512^3$
grid-zones over the same unit lengthscale. As a result, model \texttt{A}
is subsonic, model \texttt{B} is transonic and models \texttt{C} and
\texttt{D} are instead supersonic. The spatial and time-averaged
relativistic Mach numbers $\langle v W \rangle/(c_s W_s)$ are $0.362$,
$0.543$, $1.003$ and $1.759$ for our models \texttt{A}, \texttt{B},
\texttt{C} and \texttt{D}, while the average Lorentz factors are $1.038$,
$1.085$, $1.278$ and $1.732$ respectively

The initial conditions are simple: a constant energy density and a
zero-velocity field. The forcing term, which is enabled at time $t = 0$,
quickly accelerates the fluid, which becomes turbulent. By the time when
we start to sample the data, \ie at $t = 10$ (light-)crossing times,
turbulence is fully developed and the flow has reached a stationary
state. The evolution is then carried out up to time $t = 40$, thus
providing data for 15, equally-spaced timeslices over $30$ crossing
times. As a representative indicator of the dynamics of the system, we
show in the left panel of Fig.~\ref{fig:lorentz} the time evolution of
the average Lorentz factor for the different models considered. Note that
the Lorentz factor grows very rapidly during the first few crossing times
and then settles to a quasi-stationary evolution. Furthermore, the
average grows nonlinearly with the increase of the driving term, going
from $\langle W \rangle \simeq 1.04$ for the subsonic model \texttt{A},
up to $\langle W \rangle \simeq 1.73$ for the most supersonic model
\texttt{D}.

The probability distribution functions (PDFs) of the Lorentz factor are
shown in the right panel of Fig. \ref{fig:lorentz} for the different
models.  Clearly, as the forcing is increased, the distribution widens,
reaching Lorentz factors as large as $W\simeq 40$ (\ie to speeds $v\simeq
0.9997$). Even in the most ``classical'' case \texttt{A}, the flow shows
patches of fluid moving at ultrarelativistic speeds. Also shown in Fig.
\ref{fig:lorentz} is the logarithm of the Lorentz factor on the $(y,z)$
plane and at $t=40$ for model \texttt{D}, highlighting the large spatial
variations of $W$ and the formation of front-like structures.
\section{Universality}
As customary in studies of turbulence, we have analyzed the power
spectrum of the velocity field 
\begin{equation}
E_{\boldsymbol{v}}(k) \equiv \frac{1}{2} \int_{|\boldsymbol{k}|=k}
| \hat{\boldsymbol{v}}(\boldsymbol{k}) |^2\, d\boldsymbol{k}\,,
\end{equation}
where $\boldsymbol{k}$ is a wavenumber three-vector and
\begin{equation}
\hat{\boldsymbol{v}}(\boldsymbol{k}) \equiv 
\int_V \boldsymbol{v}(\boldsymbol{x})
e^{-2 \pi i \boldsymbol{k}\cdot\boldsymbol{x}}\, d\boldsymbol{x}\,,
\end{equation}
with $V$ being the three-volume of our computational domain. A number
of recent studies have analyzed the scaling of the velocity power
spectrum in the inertial range, that is, in the range in wavenumbers
between the lengthscale of the problem and the scale at which
dissipation dominates. More specifically, \citet{Inoue2011} has
reported evidences of a Kolmogorov $k^{-5/3}$ scaling in a
freely-decaying MHD turbulence, but has not provided a systematic
convergence study of the spectrum. Evidences for a $k^{-5/3}$ scaling
were also found by \citet{Zhang09}, in the case of the
kinetic-energy spectrum, which coincides with the velocity
power-spectrum in the incompressible case. Finally,
\citet{Zrake2011a} has performed a significantly more systematic
study for driven, transonic, MHD turbulence, but obtained only a very
small (if any) coverage of the inertial range.

\begin{figure}
  \begin{center}
    \includegraphics[width=0.8\columnwidth]{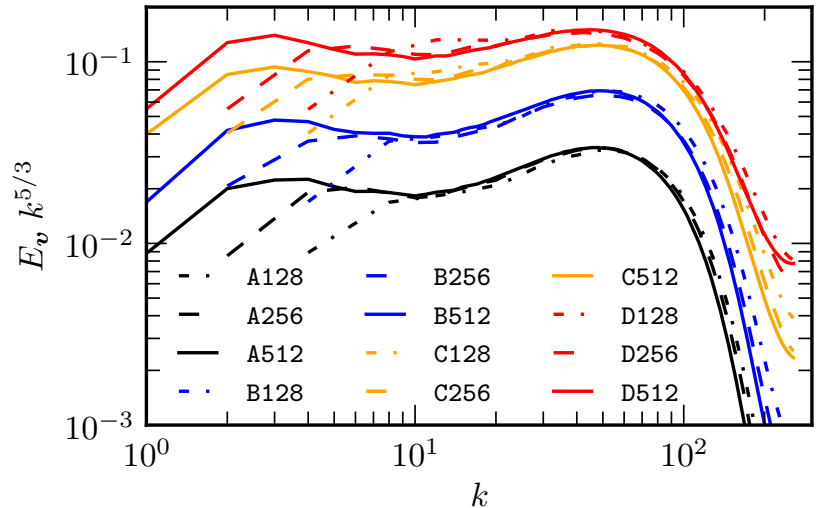}
  \end{center}
  \caption{Power spectra of the velocity field.  Different lines refer
    to the three resolutions used and to the different values of the
    driving force. The spectra are scaled assuming a $k^{-5/3}$ law.}
\label{fig:vel_spectrum}
\end{figure}

The time-averaged velocity power spectra computed from our simulations
are shown in Fig.~\ref{fig:vel_spectrum}. Different lines refer to the
three different resolutions used, $128^3$ (dash-dotted), $256^3$
(dashed) and $512^3$ (solid lines), and to the different values of the
driving force. To highlight the presence and extension of the inertial
range, the spectra are scaled assuming a $k^{-5/3}$ law, with curves
at different resolutions shifted of a factor two or four, and nicely
overlapping with the high-resolution one in the dissipation
region. Overall, Fig.~\ref{fig:vel_spectrum} convincingly demonstrates
the good statistical convergence of our code and gives a strong
support to the idea that the \emph{key} prediction of the Kolmogorov
model (K41) \citep{Kolmogorov1991a} carries over to the relativistic
case. Indeed, not only does the velocity spectrum for our subsonic
model \texttt{A} shows a region, of about a decade in length, where
the $k^{-5/3}$ scaling holds, but this continues to be the case even
as we increase the forcing and enter the regime of relativistic
supersonic turbulence with model \texttt{D}. In this transition, the
velocity spectrum in the inertial range, the range of lengthscales
where the flow is scale-invariant, is simply ``shifted upwards'' in a
self-similar way, with a progressive flattening of the bottleneck
region, the bump in the spectrum due to the non-linear dissipation
introduced by our numerical scheme. Steeper scalings, such as the
Burger one, $k^{-2}$, are also clearly incompatible with our
data. 

All in all, this is one of our main results: the velocity power
spectrum in the inertial range is \emph{universal}, that is,
insensitive to relativistic effects, at least in the subsonic and
mildly supersonic cases. Note that this does \emph{not} mean that
relativistic effects are absent or can be neglected when modelling
relativistic turbulent flows.
\section{Intermittency}
Not all of the information about relativistic turbulent flows is
contained in the velocity power spectrum. Particularly important in a
relativistic context is the intermittency of the velocity field, that
is, the local appearance of anomalous, short-lived flow features,
which we have studied by looking at the parallel-structure functions
of order $p$
\begin{equation}\label{eq:structure.function}
  S^\parallel_p(r) \equiv \big\langle |\delta_r v|^p \big\rangle,
  \qquad
  \delta_r v = \big[
    \boldsymbol{v}(\boldsymbol{x}+\boldsymbol{r}) -
    \boldsymbol{v}(\boldsymbol{x})\big] \cdot \frac{\boldsymbol{r}}{r}
\end{equation}
where $\boldsymbol{r}$ is a vector of length $r$ and the average is
over space and time.

\begin{figure}
  \begin{minipage}{0.49\hsize}
    \center{\qquad Model A}
    \includegraphics[width=\columnwidth]{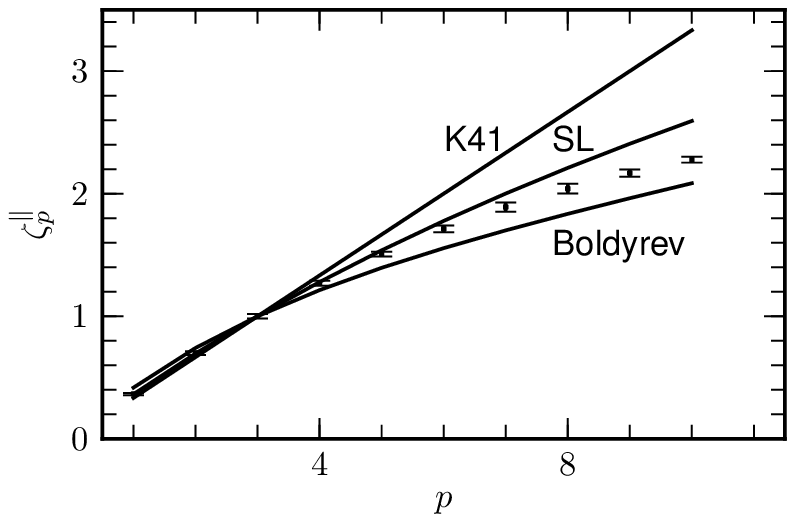}
    \vskip 0.5cm
    \center{\qquad Model C}
    \includegraphics[width=\columnwidth]{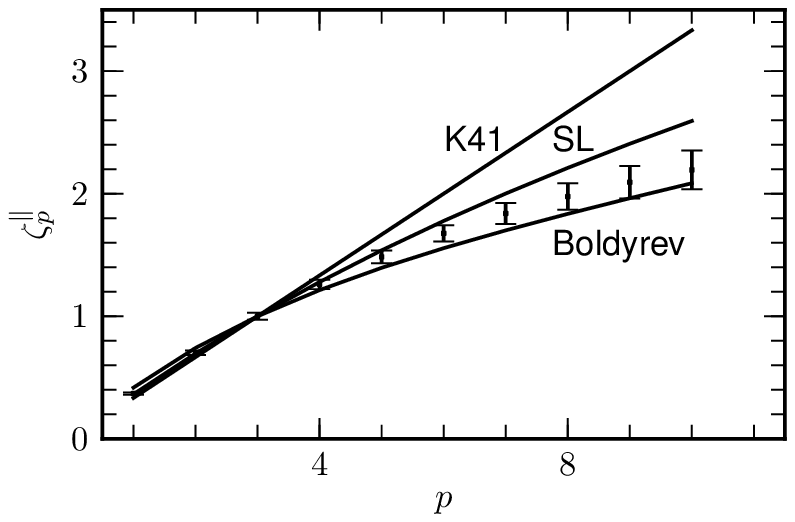}
  \end{minipage}
  \begin{minipage}{0.49\hsize}
  \begin{center}
    \center{\qquad Model B}
    \includegraphics[width=\columnwidth]{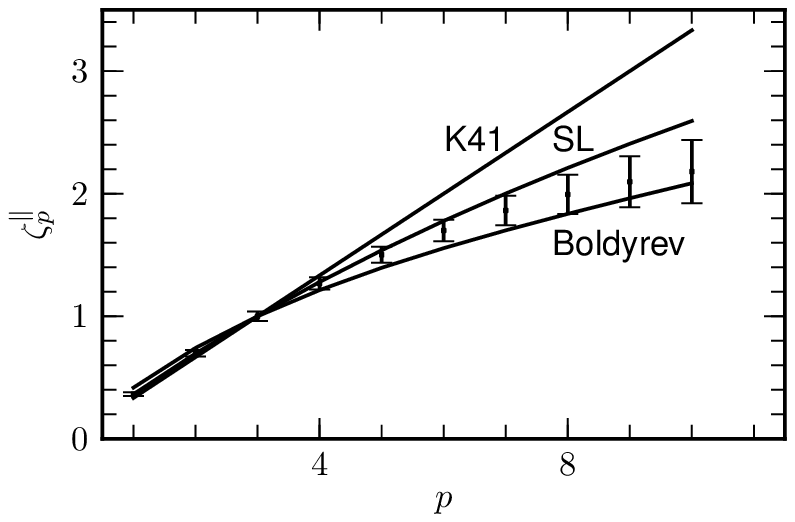}
    \vskip 0.5cm
    \center{\qquad Model D}
    \includegraphics[width=\columnwidth]{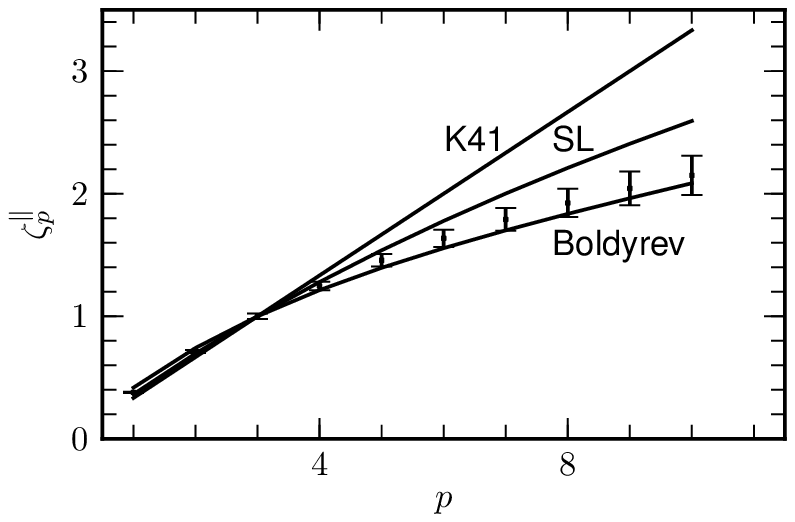}
  \end{center}
  \end{minipage}
  \caption{Structure function exponents as computed using the ESS
  technique for different models. Also shown in the Figure are the
  analytic predictions from two classical intermittency models: the She
  and Leveque (SL) \citep{She1994} and the Boldyrev \citep{Boldyrev2002}
  models.\label{fig:structure}}
\end{figure}

The scaling exponents of the parallel structure functions, \ie
$\zeta^\parallel_p$ \st $S_p^\parallel(r) \sim r^{\zeta^\parallel_p}$,
have been computed up to $p=10$ using the extended-self-similarity (ESS)
technique \citep{Benzi1993} and are summarized in
Figure~\ref{fig:structure}. The errors are estimated by computing the
exponents without the ESS or using only the data at the final time. We
also show the values as computed using the classical K41 theory, as well
as using the estimates by She and Leveque (SL) \citep{She1994} for
incompressible turbulence, \ie $\zeta^\parallel_p = \frac{p}{9} + 2 - 2
(\frac{2}{3})^{p/3}$, and those by Boldyrev \citep{Boldyrev2002} for
Kolmogorov-Burgers supersonic turbulence, \ie $\zeta^\parallel_p =
\frac{p}{9} + 1 - (\frac{1}{3})^{p/3}$.

Not surprisingly, as the flow becomes supersonic, the high-order
exponents tend to flatten out and be compatible with the Boldyrev
scaling, as the most singular velocity structures become
two-dimensional shock waves. $\zeta_2^\parallel$, instead, is
compatible with the She-Leveque model even in the supersonic
case. This is consistent with the observed scaling of the velocity
power spectrum, which presents only small intermittency corrections to
the $k^{-5/3}$ scaling. Previous classical studies of weakly
compressible \citep{Benzi2008} and supersonic
turbulence \citep{Porter2002} found the scaling exponents to be in very
good agreement with the ones of the incompressible case and to be well
described by the SL model. This is very different from what we observe
even in our subsonic model \texttt{A}, in which the exponents are
significantly flatter than in the SL model, suggesting a stronger
intermittency correction. This deviation is another important result
of our simulations.

\section{Conclusions}
Using a series of high-order direct numerical simulations of driven
relativistic turbulence in a hot plasma, we have explored the
statistical properties of relativistic turbulent flows with average
Mach numbers ranging from $0.4$ to $1.7$ and average Lorentz factors
up to $1.7$. We have found that relativistic effects enhance
significantly the intermittency of the flow and affect the high-order
statistics of the velocity field. Nevertheless, the low-order
statistics appear to be universal, \ie independent from the Lorentz
factor, and in good agreement with the classical Kolmogorov theory.

\acknowledgements
We thank M.A. Aloy, P. Cerd\'a-Dur\'an, 
and M. Obergaulinger
for discussions. The calculations were performed on the clusters at the
AEI and on the SuperMUC cluster at the LRZ. Partial support comes from
the DFG grant SFB/Trans-regio~7 and by ``CompStar'', a Research
Networking Programme of the ESF.

\bibliography{author}

\end{document}